# Stealth Non-standard-model Confined Flare Eruptions: Sudden Reconnection Events in Ostensibly Inert Magnetic Arches from Sunspots

Ronald L. Moore[1,2], Sanjiv K. Tiwari[3,4], Navdeep K. Panesar[3,4], V. Aparna[3,4], & Alphonse C. Sterling[2]

[1]Center for Space Plasma and Aeronomic Research (CSPAR), UAH, Huntsville AL, 35805 USA; ronald.l.moore@nasa.gov

[2]NASA Marshall Space Flight Center, Huntsville, AL 35812, USA

[3]Lockheed Martin Solar Astrophysics Laboratory, 3251 Hanover Street Building 203, Palo Alto, CA 94306, USA; tiwari@lmsal.com

[4]Bay Area Environmental Research Institute, NASA Research Park, Moffett Field, CA 94035, USA

## Abstract

We report seven examples of a long-ignored type of confined solar flare eruption that does not fit the standard model for confined flare eruptions. Because they are confined eruptions, do not fit the standard model, and unexpectedly erupt in ostensibly inert magnetic arches, we have named them stealth non-standard-model confined flare eruptions. Each of our flaring magnetic arches stems from a big sunspot. We tracked each eruption in full-cadence UV and EUV images from the Atmospheric Imaging Assembly (AIA) of Solar Dynamics Observatory (SDO) in combination with magnetograms from SDO's Helioseismic and Magnetic Imager (HMI). We present the onset and evolution of two eruptions in detail: one of six that each make two side-by-side main flare loops, and one that makes two crossed main flare loops. For these two cases, we present cartoons of the proposed pre-eruption field configuration and how sudden reconnection makes the flare ribbons and flare loops. Each of the seven eruptions is consistent with being made by sudden reconnection at an interface between two internal field strands of the magnetic arch, where they cross at a small (10° - 20°) angle. These stealth non-standard-model confined flare eruptions therefore plausibly support the idea of E. N. Parker for coronal heating in solar coronal magnetic loops by nanoflare bursts of reconnection at interfaces of internal field strands that cross at angles of 10° - 20°.

## 1. Introduction

This paper is dedicated to Frances Tang. Ron Moore was a colleague of Frances at Caltech in the 1970's, when he was a postdoc in the Solar Astronomy Group of Hal Zirin's Big Bear Solar Observatory (BBSO). Frances brought to Ron's attention "umbral" flares that she found in searching BBS0's Hα movies of solar active regions. Those flares are germane to the work reported here.

In this paper, the term "flare eruption" has a broader meaning than the term "eruptive flare." "Eruptive flare" commonly refers to any flare that is made by a solar magnetic explosion that becomes a coronal mass ejection (CME) (e.g., Svestka et al 1992). The term "confined flare" commonly refers to any flare that is made by sudden magnetic energy release that is confined within enveloping magnetic field that prevents the exploding field from becoming a CME. In this paper, following Moore & Roumeliotis (1992) and Moore et al (2001), we use the term "ejective flare" instead of "eruptive flare" for any flare made in tandem with a CME. We use the term "flare eruption" as a general term for the process of sudden release of magnetic energy from the magnetic field in which a flare occurs. Any flare's bright emissions of all wavelengths presumably mostly result from plasma particle energization via sudden reconnection of the flare-making magnetic field. In our usage, the term "flare eruption" stands for the sudden magnetic energy release process and resulting flare production in any type of flare, ejective or confined.

Solar flare and flare-like eruptions of all sizes – from big ones that become fast wide CMEs, to comparatively small ones that make X-ray and EUV coronal jets, to yet smaller ones seen as EUV jetlets and campfires in magnetic network, EUV transient bright dots in emerging and recently emerged active regions, and UV penumbral jets in sunspots – are evidently produced by explosive release of energy from the magnetic field in which the eruption occurs, undoubtedly via reconnection of that field (e.g., Sturrock 1980; Machado et al 1988; Moore et al 1999, 2001, 2011; Innes & Teriaca 2013; Priest 2014; Sterling et al 2015; Panesar et al 2016, 2018, 2021; Tiwari et al 2016, 2018, 2019, 2022; Raouafi et al 2023). For some flare eruptions, the pre-eruption magnetic field is a closed bipole that sits far from other field of comparable strength. That is the case for each of the six flare eruptions shown in Moore et al (2001). Most such single-bipole flare eruptions fit the so-called standard model for single-bipole flare eruptions that is sketched in Figure 1 and is borne out by recent MHD simulations (Moore et al. 2001; Jiang et al. 2021, 2023; Zhong et al. 2023).

A lone bipolar magnetic field that is set to undergo a flare eruption typically has the following three-dimensional form evident from photospheric magnetograms in combination with chromospheric and coronal images. The bipole's two opposite-polarity photospheric magnetic flux domains sit close against each other so that the polarity inversion line (PIL) that runs through the core of the bipole is fairly sharp. The core field, that is, field rooted near the PIL, is greatly sheared so that it is directed nearly along the PIL, instead of nearly orthogonally to the PIL as it would be if the core field were nearly in its zero-free-energy potential-field configuration. That shows the core field has lots of free magnetic energy, some of which might possibly be released in a flare eruption. In the chromosphere and low corona, the sheared core field is seen as a so-called filament channel, a stream of fine-scale chromospheric fibrils directed along the sheared core field. The core field also often suspends within it a long dark filament of chromospheric-temperature plasma. The filament is suspended above the channel's fine-scale fibrils and is thicker than a single fibril. In coronal images, the core field has the overall sigmoidal form of either the letter S or the numeral 2 (backward S). The middle of the sigmoid traces the PIL through the core of the bipole. Much of the magnetic field in the middle of the sigmoid is in the two opposite-polarity forearms of two oppositely curved elbows that together fill much of the whole sigmoid. The two forearms shear past each other in the core of the bipole. There is some “end-to-end” field in the sigmoid, field that is not rooted in the middle of the bipole as the forearms are but is rooted only in the two outer ends of the sigmoid. The end-to-end field is basically a mildly twisted sigmoidal flux rope that dips through the core of the bipole and runs along and above the PIL there. The filament sits in the dip. In the middle of the sigmoid, the filament-holding end-to-end field, is laterally narrowly sandwiched between the forearms of the sigmoid’s two elbows. The end of each elbow’s forearm is in flux of one polarity near the PIL in the core of the bipole, and the far end of that elbow is in the bipole’s outer flux of the opposite polarity, as in Figure 1. The bipole’s magnetic field enveloping the core field is a magnetic arcade that arches increasingly more orthogonally over the PIL with increasing distance from the PIL. A classic example of a pre-flare-eruption lone bipolar magnetic field that displays the above 3D morphology is shown in Moore & LaBonte (1980).

Before the start of reconnection in the standard model, the sigmoid’s two opposite polarity elbow forearms are only narrowly separated by the sigmoid’s end-to-end field, and are thus poised to start reconnecting as soon as they come sufficiently into contact by pushing through the intervening end-to-end field, perhaps as a consequence of convection-driven magnetic flux cancellation at the PIL, as suggested by van Ballegooijen & Martens (1989) and Moore & Roumeliotis (1992). Because the reconnection starts in the chromosphere or low corona, the two opposite-polarity flare ribbons heated at the opposite feet of the downward reconnected field lines are at first short and closely bracket the PIL under the middle of the sigmoid and filament as the filament begins to erupt upward (as in the flare eruption presented in Moore & LaBonte 1980). As the eruption of the filament and filament-carrying flux rope accelerates, the vertical and horizontal extents of the reconnection interface between the two forearms rapidly grow, as the flare ribbons rapidly lengthen along the PIL and start spreading apart.

There are two alternatives for how the ensuing flare eruption plays out: the eruption is either confined or ejective. That is, for a flare eruption as large as a big solar active region or larger (spanning $>\sim 10^5$ km), the eruption either does not become a CME because it is confined or does become a CME because it is ejective. In a single-bipole confined flare eruption that fits the standard model, the eruption of the filament-carrying flux rope is arrested within bipole’s closed field that envelops the sheared core field. The opposite flare ribbons start close to the PIL and spread apart little if any, e.g., as in the flare eruption presented in Moore (1988). In a standard-model ejective flare eruption, the filament/flux-rope eruption blows open the bipole’s envelope and normally becomes a CME if the bipole is the size of a big active region or larger. As in Figure 1, the flare ribbons last much longer and spread apart much more than in a confined flare eruption.

It has long been observed that there is a small minority of single-bipole confined flare eruptions that do not fit the standard model. Tang (1978) presents five such flare eruptions observed in chromospheric Hα images. Each of these is appropriately called an umbral flare. Each flare erupts in a seprate lone bipolar active region. Each active region has a single large sunspot in its leading magnetic polarity domain, and has in its following polarity domain either only plage or both plage and smaller sunspots. Contrary to the setup for single-bipole confined flare eruptions that fit the standard model, the PIL of each active region’s bipolar magnetic arch is not traced by a filament channel and filament. Instead, the PIL is roughly orthogonally crossed by arched filaments connecting the magnetic arch’s two opposite-polarity domains. That indicates the magnetic arch’s field rooted near the PIL is not greatly sheared across the PIL, and so has no obvious large store of free magnetic energy such as the pre-eruption sheared core field has in the standard model. As the flare ribbons brighten, none of the arched filaments erupt and none are disturbed. Each flare eruption has only one pair of ribbons, and the two ribbons are widely separated. One ribbon sits in the umbra of the large leading sunspot and is more compact than the other ribbon. The other ribbon sits in the middle of following-polarity plage or plage and sunspots. The two ribbons suddenly brighten in place simultaneously and grow to their maximum brightness and area with little or no spreading apart. From the placement and growth of both ribbons, we infer the flare energy release is by reconnection that suddenly starts between entwined strands of magnetic field that arches high over the PIL. That burst of reconnection heats the magnetic arch's interior field that is rooted in and connects the two ribbons. Thus, each of these five single-bipole confined flare eruptions obviously does not fit the standard model shown in Figure 1.

Tandberg-Hanssen et al. (1984) present another single-bipole flare eruption that appears to be a confined eruption and that does not fit the standard model. That flare eruption is in a bipolar magnetic arch that is part of a bipolar active region. The flare erupts in the same manner as the single-bipole flare eruptions of Tang (1978), but the form of the magnetic field rooted near the PIL of the magnetic arch is different. The PIL is traced by a filament in Hα images. That shows the core field along the PIL under the magnetic arch is strongly sheared and hence has a large store of free energy, more or less like the pre-eruption core field in the standard model. However, the filament neither erupts nor is disturbed during the flare eruption. This indicates that, contrary to the standard model, the magnetic energy release for this flare does not involve the sheared core field. Another difference from the single-bipole flare eruptions of Tang (1978) is that neither of the two flare ribbons is in a sunspot. The opposite ribbons are in opposite-polarity plage without sunspots. Even so, the flare eruption is similar to those of Tang (1978) in that the two flare ribbons are each offset from the magnetic arch's PIL traced by the filament between them, and suddenly brighten and grow in place with little or no spreading apart. Tandberg-Hanssen et al. (1984) therefore infer – as we have inferred for the single-bipole flare eruptions of Tang (1978) – that the flare energy release is by sudden reconnection of entwined strands of the magnetic arch's interior magnetic field that arches high over the PIL and is rooted in and connects the two flare ribbons. Thus, in the same way as the single-bipole confined flare eruptions of Tang (1978), the single-bipole flare eruption presented by Tandberg-Hanssen et al. (1984) obviously does not fit the standard model for confined single-bipole flare eruptions shown in Figure 1.

This paper reports our investigation of seven single-bipole confined flare eruptions that, in the manner of the ones reported by Tang (1978) and Tandberg-Hanssen et al. (1984), do not fit the standard model for single-bipole confined flare eruptions. Each of our seven eruptions is well observed with good spatial and temporal resolution by UV and EUV images from the Atmospheric Imaging Assembly (AIA, Lemen et al. 2012) of Solar Dynamics Observatory (SDO, Pesnell et al. 2012) in combination with magnetograms from SDO's Helioseismic and Magnetic Imager (HMI, Scherrer et al. 2012). Each eruption suddenly starts in the interior of a bipolar magnetic arch that has one foot in a large sunspot and the other foot in plage of the opposite magnetic polarity. The magnetic arch shows no pre-eruption evidence of its internal magnetic structure that is poised to unleash the flare eruption via reconnection. From the observed flare loops and other observed aspects, we characterize the development and configuration of each flare eruption's main heated magnetic loops. The seven eruptions have two different arrangements of main flare loops. For each of the two flare-loop arrangements, we present an example eruption in detail, and propose the three-dimensional form of the pre-eruption magnetic field that reconnects in the eruption.

## 2. Data and Methods

The main data are AIA images and HMI magnetograms from SDO. In addition, we used the web site solarmonitor.org to get (1) the NOAA Active Region Number of the active region of each of the active regions in which our seven flare eruptions occur, and (2) time plots of 5-minute averages of the Sun's 1 – 8 Å soft X-ray flux from NOAA GOES satellites to get the GOES X-ray magnitude of each flare.

The AIA field of view is square, normally centers on the Sun, and spans the entire solar disk and low corona. The image pixels are square, each 0.6 arc second wide. AIA has two UV channels and seven EUV channels. The cadence of the images from each UV channel is 24 s. The cadence of the images from each EUV channel is 12 s. We mainly use images from the UV channel that is centered at 1600 Å and images from four of the EUV channels, the ones centered at 304 Å, 171 Å, 211 Å, and 131 Å. The 1600 Å channel images UV continuum from the upper photosphere not in flares, and C IV emission from flare-heated plasma at temperatures around $10^5$ K. The 304 Å channel images He II emission from plasma at cooler-transition-region temperatures around 5 x $10^4$ K. The 171 Å channel images Fe IX emission from plasma at hotter-transition-region temperatures around 6 x $10^5$ K. The 211 Å channel images Fe XIV emission from plasma at coronal temperatures around 2 x $10^6$ K. The 131 Å channel images both Fe VIII emission from plasma at transition-region temperatures around 4 x $10^5$ K and Fe XXI emission from flare-heated plasma at temperatures around $10^7$ K.

The images from the 304 Å and 171 Å channels best show whether the PIL of the magnetic arch in which the flare eruption occurs is traced by a filament channel and/or a dark filament before flare eruption onset.

Among the images from AIA's nine channels, the onset and growth of each eruption's flare ribbons at the feet of the reconnection-heated loops are most visible in full-cadence (24 s) 1600 Å images and full-cadence (12 s) 131 Å images. In the 131 Å images, the brightening of the feet of a flare loop leads the brightening of the whole body of the loop by several time steps. This is consistent with the 131 Å channel being roughly equally sensitive to emission from plasma at temperatures around 4 x $10^5$ K as it is to emission from plasma at temperatures around $10^7$ K, as follows. Plausibly, the temperature of the coronal plasma in the body of a reconnected flare loop is initially (immediately after reconnection) somewhat (no more than about twice) hotter than $10^7$ K, but that hot plasma's density is initially not enough for the loop body to be bright in AIA 131 Å images. The reconnection-heated coronal plasma in the flare loop body then increases in density and decreases in temperature via so-called "chromospheric evaporation" from the loop's feet as a result of sudden strong heating of chromospheric-temperature dense plasma in the loop's feet by non-thermal particles from the reconnection and by heat conduction from the initially tenuous hot plasma in the loop body. Loop-foot plasma is thereby suddenly heated to temperatures of $\sim 10^7$ K and therefore expands up into the body of the flare loop. When the loop-body plasma becomes dense enough, the chromospheric evaporation stops, the loop-body plasma density stops increasing, and the loop-body plasma continues to cool by radiation and by heat conduction to the loop feet. Hence, first, a flare-loop-foot brightening turns on simultaneously in 1600 Å and 131 Å AIA images, leading the brightening of the loop body in the 131 Å images. Then, the loop body brightens and fades in the 131 Å images before it brightens and fades in the 211 Å images and soon after in the 171 Å images. Among the seven AIA **EUV** channels, this progressively delayed brightening in a flare loop's images from progressively cooler channels is most obvious from the loop's peaking in brightness first in 131 Å images, later in 211 Å images, and still later in 171 Å images.

We use HMI line-of-sight photospheric magnetograms. The HMI field of view is square, normally centers on the Sun, and is somewhat wider than the solar disk. The line-of-sight magnetograms have 0.5 arc second square pixels and 45 s cadence. Their noise level is about 10 G (Schou et al. 2012; Couvidat et al. 2016). For each of our selected flare eruptions, the HMI magnetograms, together with co-aligned AIA images of the eruption's heated flare loops and the heated flare ribbons at the flare loop feet, show the strength and polarity of the magnetic flux at each flare loop foot. These also show whether the magnetic flux in the flare ribbon at a flare loop's foot is purely unipolar, i. e., whether any minority-polarity flux is discernible there in the magnetograms.

To find our stealth non-standard-model single-bipole confined flare eruptions, we used Helioviewer to search through three years of AIA images and co-temporal HMI magnetograms, from the start of 2012 to the end of 2014. We looked for bipolar flare eruptions that are similar to those of the umbral flares of Tang (1978). That is, we looked for two-ribbon flares for which (1) the eruption is rooted in a bipolar magnetic arch that has one end in unipolar umbra and/or penumbra of a large sunspot and the other end in unipolar plage of the opposite polarity, (2) the two opposite-polarity flux domains in which the arch is rooted are not tightly packed against each other at the PIL between them under the arch, i.e., the PIL is not sharp but vaguely defined in the magnetograms, (3) little or none of the PIL under the arch is traced by a filament channel or dark filament seen in 304 Å and 171 Å AIA images, (4) there is a single flare ribbon in each foot of the arch, (5) each ribbon starts well away from the PIL, and grows bigger and brighter and dims around where it starts, with little or no spreading apart of the two ribbons, (6) the PIL-tracing filament and/or filament channel, if any, neither participates in nor is disturbed by the flare eruption, and (7) in the pre-eruption AIA images and HMI magnetograms there is no sign of magnetic structure that is set to suddenly reconnect in the body of the arch; the pre-eruption magnetic arch is ostensibly inert. Our search yielded seven two-ribbon flare eruptions that meet these selection criteria.

In the above respects, each selected flare eruption is a single-bipole flare eruption that is similar to the umbral flares of Tang (1978). In the same manner as those, each selected eruption obviously does not fit the standard model for single-bipole confined flare eruptions shown in Figure 1. Our selected seven non-standard-model single-bipole confined flare eruptions are listed in Table 1.

For each of the flare eruptions in Table 1, we made a full-cadence HMI magnetogram movie and full-cadence AIA 1600 Å, 304 Å, 171 Å, 211 Å, and 131 Å movies. The field of view of each eruption's movies is centered on and zoomed in on the area of the eruption. Each movie starts minutes before flare onset and ends far into the decay of the flare's coronal EUV luminosity. We used these to track the onset and progression of the flare ribbons and loops and their arrangement in each flare eruption.

## 3. Results

### *3.1. Table 1*

Table 1 lists our main empirical results. The chronological number of each of the seven eruptions is in the first column. The eruption's date of occurrence and heliographic location are in the second column, and the NOAA number of the eruption's active region is in the third column. Whether the eruption made a flare or only a subflare, and the magnitude of the eruption's GOES X-ray burst, are given in the fourth column. In that column, the term Flare is for any eruption having GOES magnitude $\geq$ C1, and the term Subflare is for any eruption having GOES magnitude < C1.

The fifth and sixth columns give our assessment of the sharpness the magnetic field's polarity reversal across the PIL of the pre-eruption magnetic arch and of whether the core field closely enveloping the PIL shows evidence of being strongly sheared via tracing of the PIL by a filament or filament channel. The fifth column gives our Yes/No judgement of whether the PIL appears sharp in the magnetograms. For that, our judgement is No for each of the seven eruptions. The sixth column gives our Yes/No judgement of whether any of the pre-eruption PIL and core field are traced by a filament and/or filament channel in AIA 304 Å and 171 Å images. For that, our judgement is Yes for only eruptions 1 and 4, and is No for each of the other eruptions.

The seventh through tenth columns concern each eruption's heated flare loops. The seventh and eighth columns give a start time and an end time for each eruption's flare loops and ribbons. The start time is the time of the first discernible brightening of the flare ribbons in AIA 131 Å and 1600 Å images. The listed end time is a time late in the decay of the brightness of the loops in AIA 171 Å images.

The ninth column characterizes the configuration of the eruption's final main flare loops seen in AIA images. Of our seven flare eruptions, there are two different categories of the configuration of the final main flare loops: two side-by-side loops and two crossed loops. The configuration is two side-by-side loops for six of our seven eruptions (all but eruption 1). Only in eruption 1 is the configuration two crossed loops.

The last column of Table 1, the tenth column, notes aspects of each flare eruption and its heated flare loops and ribbons that do not fit the standard model for singe-bipole confined flare eruptions. In all seven eruptions, the core field along the PIL of each flaring magnetic arch is inert, the flare ribbons start far from the PIL, and the flare ribbons do not noticeably spread apart.

### *3.2. Example Eruptions*

#### *3.2.1. An Eruption that Makes Two Side-by-Side Flare Loops*

Eruption 2 is representative of the six eruptions (eruptions 2, 3, 4, 5, 6, 7) in which the final flare-loop configuration is two side-by-side loops.

Figure 2 is one frame from Figure 2 animation. It is near the time of maximum luminosity of eruption 2 in AIA 131 Å emission. The AIA 131 Å image in Figure 2 shows the eruption 2 flare loops in their final configuration. The AIA 1600 Å, 211 Å, and 171 Å images each show the flare ribbons at the loop feet. The 1600 Å image, the magnetogram, and the 1600 Å image overlaid with magnetogram contours, together show the magnetic setting of the flare loops and ribbons. The animation covers the 55-minute duration of eruption 2 given in Table 1. It shows at full cadence the progression of eruption 2 spanned by the four time steps in Figure 3.

The magnetograms and images in Figure 3 are centered on eruption 2 and step through the progression of the eruption's flare ribbons and loops. They all have the same field of view. In each row, the five panels are at nearly the same time. From left to right they are an HMI magnetogram, an AIA 1600 Å image, an AIA 131 Å image, an AIA 211 Å image, and an AIA 171 Å image. In the top row, the 20 G and 100 G contours of the magnetogram (panel a) are superposed on the 1600 Å image (panel b) . The time of the top row is about a minute before the first discernible brightening of the flare ribbons in 1600 Å, 131 Å, 211 Å, and 171 Å images in Figure 2 animation. The time of the second row is about 5 minutes later, in the onset of the flare ribbons in 1600 Å, 211 Å, and 171 Å images and in the onset of the flare ribbons and loops in 131 Å images. The time of the third row is another 4 minutes later, during the fastest rise of the flare's luminosity in 131 Å images. The time of the fourth row is yet another 4 minutes later, at the end of the rise of the flare's luminosity in 131 Å images. The time in the bottom row is 19 minutes later than in the fourth row in panels u through x and 4 more minutes later in panel y (171 Å). The images in panels u through y are from well after flare loop heating has ended. They are at times when the flare loops have cooled to become much dimmer in the 131 Å image, and the flare ribbons have dimmed to near invisibility in the 1600 Å image.

From comparison of the magnetograms in Figure 3 with the flare ribbons and loops in the 1600 Å and 131 Å images, this flare eruption is evidently in a north-south magnetic arch that has its northern foot in the umbra and penumbra of the largest negative-polarity sunspot of AR 11654 and has its southern foot in positive-polarity flux south of the that sunspot. The magnetograms show that the east-west PIL bridged by the magnetic arch is ragged. In the 171 Å images, no east-west filament or filament channel is seen under the arch. Instead, there are only some north-south dark fibrils across the east-west PIL. In the 171 Å images in Figure 3 and in the 171 Å movie in the Figure 2 animation, these fibrils hardly change during the flare eruption. The flare ribbons at opposite ends of the flare loops start far from the PIL and do not spread apart as they grow bigger and brighter. There is no sign that the flare eruption blows open the magnetic arch in which is erupts. Thus, this flare is evidently a confined eruption that evidently does not fit the standard model for single-bipole confined flare eruptions shown in Figure 1.

In the 1600 Å images in Figure 3, the opposite-polarity conjugate pair of flare ribbons first appear in the second image (panel g). In the third image (panel l), each ribbon has grown bigger and brighter in place, without noticeably spreading away from the other ribbon. Both ribbons have noticeably dimmed in place in the fourth image (panel q), and have nearly faded out in the fifth image (panel v). In the second, third and fourth 1600 Å images (panels g, l, q), each opposite polarity ribbon has two side-by-side parts: a western part and an eastern part. In each of these three images, a white arrow points to the western part of the negative-polarity northern ribbon, and another white arrow points to the western part of the positive-polarity southern ribbon. A black arrow points to the eastern part of the northern ribbon. The eastern part of the southern ribbon itself has two parts: a northern part and a southern part. The two prongs of a black bracket at the head of a black arrow each point to one of the two parts of the eastern part of the southern ribbon. The white arrows, black arrows, and black bracket are repeated in the second, third, and fourth images in the 131 Å, 211 Å, and 171 Å columns of Figure 3, each pointing to the same ribbon part it points to in the corresponding 1600 Å image.

In the second, third, and fourth rows of Figure 3, the flare ribbon at either end of the flare loops is at the same place and shows the same eastern and western parts in all four AIA channels. Evidently the loop feet have brightly emitting plasma ranging in temperature from as at least as cool as $10^5$ K (for the 1600 Å ribbons) up through 6 x $10^5$ K (for the 171 Å ribbons) and 2 x $10^6$ K (for the 211 Å ribbons) to perhaps as hot as $10^7$ K (for the 131 Å ribbons). During the time the flare ribbons are obvious in Figure 3, the flare loops – from which undoubtedly comes the heating of the ribbons – are visible in the 131 Å images but not in the 1600 Å, 211 Å, and 171 Å images. Evidently, during that time, except at the loop feet, the plasma in the flare loops is still too hot to be seen in these cooler AIA channels. After the loop heating ends, the loop-body plasma cools to temperatures in the range of the 211 Å and 171 Å images, and some of the loops become at least partially visible in these images, as in the fifth row of Figure 3 (panels x and y).

The shorter red arrows in Figure 3 point to the west side of the flare loops, and the longer red arrows point to point to the east side of the flare loops. In the 131 Å image in the second row of Figure 3 (panel h), the body of a flare loop early in the eruption is faintly visible. In the third row (panel m), the bodies of the heated flare loops are much brighter, and are in two main bundles that cross each other at a small angle (10° - 20°). The brighter bundle has its southern foot in the eastern part of the southern ribbon and its northern foot in the western part of the northern flare ribbon. The fainter bundle has its southern foot in the western part of the southern ribbon and its northern foot in the eastern part of the northern ribbon. In the fourth row (panel r), the 131 Å flare loops have evolved to their final configuration. There are now two side-by-side main bundles of flare loops: a western main bundle and an eastern main bundle. The western bundle is shorter and brighter and shows less substructure. The eastern bundle is longer and shows some sub-strands. The northern foot of the western bundle is the western part of the northern ribbon and the southern foot is the western part of the southern ribbon. The northern foot of the eastern bundle is the eastern part of the northern ribbon and most or all of the southern foot is the eastern part of the southern ribbon. In the fifth row, both main bundles have dimmed in 131 Å emission, evidently cooling to temperatures well below $10^7$ K, with plasma in the western bundle that has cooled to T ~ 6 x $10^5$ K being bright in the 171 Å image and also in the 131 Å image. In the fifth row, the 211 Å image (panel x) shows the western bundle (pointed to by the shorter red arrow) at near its maximum brightness in 211 Å images in Figure 2 animation. The longer red arrow in the 211 Å image of the fifth row (panel x) points to the eastern bundle, which is now starting to faintly appear in 211 Å images. In Figure 2 animation, the eastern bundle peaks in 211 Å brightness about ten minutes later, and peaks in 171 Å brightness a few minutes after that. The red arrow in the 171 Å image in the fifth row (panel y) points to the western bundle near its maximum brightness in 171 Å images.

We take the magnetic placement, spatial arrangement, and thermal evolution of the flare ribbons and loops in Figure 3 to be reasonably consistent with the final two side-by-side main flare loops having been made and heated by reconnection of two pre-flare magnetic strands that were inside the magnetic arch, crossed each other at a small (10° - 20°) angle, and suddenly reconnected at their interface.

### *3.2.2. An Eruption that Makes Two Crossed Flare Loops*

The final flare-loop configuration in eruption 1 is two crossed loops. Of our seven eruptions, eruption 1 is the only one in that flare-loop-configuration category.

Figure 4 is one frame from Figure 4 animation. It is near the time of maximum luminosity of eruption 1 in AIA 131 Å emission. The AIA 131 Å image in Figure 4 shows the eruption 1 flare loops in their final configuration. The AIA 1600 Å, 211 Å, and 171 Å images each show the flare ribbons at the loop feet. The 1600 Å image, the magnetogam, and the 1600 Å image overlaid with magnetogram contours, together show the magnetic setting of the flare loops and ribbons. The animation covers the 35-minute duration of eruption 1 given in Table 1. It shows at full cadence the progression of eruption 1 spanned by the four time steps in Figure 5.

In the same way that Figure 3 does for eruption 2, the magnetograms and images in Figure 5 show the magnetic setting and progression of the flare ribbons and loops in eruption 1. The format of Figure 5 is the same as in Figure 3. From left to right, the five columns are HMI magnetograms, AIA 1600 Å images, AIA 131 Å images, AIA 211 Å images, and AIA 171 Å images, and in the top row, the 20 G and 100 G contours of the magnetogram (panel a) are overlaid on the 1600 Å image (panel b). In Figure 5, the time of the top row is about a minute before the flare ribbons start to turn on in 1600 Å, 131 Å, 211 Å, and 171 Å images in the animation for Figure 4. The time of the second row is 3 minutes later, about 2 minutes into the brightening of the flare ribbons in 1600 Å, 211 Å, and 171 Å images and into the brightening of the flare ribbons and loops in 131 Å images. The time of the third row is another 3 minutes later, during the fastest rise of the flare's 131 Å luminosity. The time of the fourth row is yet another 4 minutes later, at the end of the rise of the flare's 131 Å luminosity. The time of the fifth (bottom) row is 7 minutes later than in the fourth row in the fifth row's first four panels (panels u, v, w, x) and another 4 minutes later in the fifth row's 171 Å panel (panel y). The time of the bottom row's first four panels is when the flare loops have cooled to their peak brightness in 211 Å images, and the time of the bottom row's 171 Å image is when the flare loops have cooled to their peak brightness in 171 Å images.

Figure 5 shows that the flare loops arch east-west, connecting the two flare ribbons. The western ribbon slants northeast-southwest, and the eastern ribbon runs roughly north-south. In the 1600 Å images, the bright southwest end of the western ribbon is seen to be in the inner penumbra of the negative-polarity largest sunspot of AR 11391. The bright northeast end of the western ribbon is in negative-polarity plage at the south edge of a nearby negative-polarity small sunspot. The eastern ribbon is in positive-polarity plage east of the big sunspot. Thus, Event 1 evidently erupts inside an east-west magnetic arch that connects that plage's magnetic flux to negative flux in and near the northeast penumbra of the big sunspot. In the magnetograms, that magnetic arch's PIL is not sharp but ragged and broken, and runs roughly east-northeast away from the big sunspot. The white arrow in the 171 Å images points to a dark filament that tracks some of that PIL. The filament remains static and shows no involvement with the flare eruption. The two flare ribbons start far from the PIL and grow bigger and brighter in place, without noticeably spreading apart. Also, there is no sign that that the flare eruption blows open the magnetic arch in which it erupts. Thus, like eruption 2, eruption 1 is evidently another confined flare eruption that evidently does not fit the standard model for single-bipole confined flare eruptions shown in Figure 1.

In each row of Figure 5, from comparison of the sunspots and plage pattern in the 1600 Å image with the corresponding features in the magnetogram, we judge that the 1600 Å image and the magnetogram are co-aligned to within a few arc-seconds (several pixels). In the second, third, and fourth rows, from close comparison of the 1600 Å image with the magnetogram, it can be seen that each of the bright ends of the western flare ribbon is definitely in negative flux. Close to where that ribbon's relatively faint and narrow middle segment, running from one bright end to the other, crosses the northeast outer edge of the big sunspot's penumbra, the magnetograms and the magnetogram contours in Figures 4 and 5 show two tiny islands of positive flux embedded in negative flux at the outer edge of the penumbra. In the magnetogram contour map in Figure 4, these two islands of positive flux are each pointed to by an arrow. Because the ribbon's middle segment curves smoothly from one bright end to the other, and from where we judge its path to be in the magnetograms, we judge that (to within the few-arc-second uncertainty in the co-alignment) the faint middle segment runs between the two positive-flux islands and is rooted in negative **flux** between the two positive-flux islands. On this basis, we judge that the two tiny positive islands do not actively participate in eruption 1, and that the western flare ribbon is all rooted in negative flux.

In the second row of Figure 5, the two flare ribbons are each early in their brightening and growth in 1600 Å and 131 Å emission (panels g and h). In the 1600 Å image and in the 131 Å image, the eastern ribbon as yet consists of only two bright points. One is near where the south end of the full-grown eastern ribbon will be, and the other is near where the north end of the full-grown ribbon will be. In both images, the southern bright point is brighter than the northern bright point, and an upward black arrow points to it. A downward black arrow points to the northern bright point. In both images, the western ribbon is brightest at its two ends. The western ribbon's northern end is brighter than its southern end, and an upward black arrow points to it. Another downward black arrow points to the western ribbon's southern end. The four black arrows are repeated in all of the AIA images in the second, third, and fourth rows of Figure 5, and each points to the same place in each image. In the second, third, and fourth rows, each ribbon is obvious in the 1600 Å and 131 Å images and is discernible in the 211 Å and 171 Å images. This indicates that the plasma temperature in the ribbons as the flare loops brighten in 131 Å emission ranges from at least as cool as $10^5$ K to perhaps as hot as $10^7$ K.

In the 131 Å image in the second row of Figure 5 (panel h), a flare loop connecting the southern bright point of the eastern flare ribbon to the bright northern end of the western flare ribbon is becoming faintly discernible. In the 131 Å image in the third row (panel m), that flare loop is now much brighter, and a fainter longer loop apparently crosses over the brighter loop at a small (10° - 20°) angle to connect the northern end of the eastern flare ribbon to the southern end of the western flare ribbon. A downward red arrow points to the fainter and longer loop, and an upward red arrow points to the brighter and shorter loop. In the 131 Å image in the fourth row (panel r), each of the two loops have brightened further, with the shorter loop still the brighter of the two. The downward red arrow again points to the fainter and longer loop, and the upward red arrow again points to the brighter and shorter loop. In the bottom (fifth) row of Figure 5, the ribbons have faded out in the 1600 Å image (panel v) but are still visible to some degree in the 131 Å, 211 Å, and 171 Å images (panels w, x, y). The red arrow in the 131 Å, 211 Å, and 171 Å images (panels w, x, y) points to the loop that earlier was the brighter and shorter of the two loops in 131 Å emission.

Similarly to the magnetic placement, spatial arrangement, and thermal evolution of the flare ribbons and loops in Figure 3 for eruption 2, the magnetic placement, spatial arrangement, and thermal evolution of the flare ribbons and loops in Figure 5 appear reasonably consistent with the two crossed main flare loops in eruption 1 having been made and heated by reconnection of magnetic-field strands that, before reconnection onset, were entwined around each other inside the magnetic arch, and then suddenly underwent a burst of reconnection.

## 4. Interpretation

As is presently well accepted, for every solar flare eruption – of any size and of any type – we assume: (1) the eruption is the manifestation of some combination of accelerated electrons and ions, heated plasma, and mass motion resulting from sudden release of energy from the magnetic field in which the eruption occurs, and (2) the magnetic field releases much of the energy via reconnection. In this Section, for each of our two example non-standard-model confined flare eruptions (eruptions 2 and 1 in Table 1), we present a schematic depicting our proposed pre-eruption magnetic field configuration and its transformation by reconnection to make the eruption's observed flare loops and ribbons. Each schematic is in the style of that in Figure 1 for the standard model for single-bipole flare eruptions. The essential aspects of the magnetic field and its reconnection are represented by a few rudimentary field lines before and after the eruption.

### *4.1.* ***Eruption*** *2*

Eruption 2, shown in Figure 3, is a good example of the six Table 1 flare eruptions for which the final flare-loop configuration is two side-by-side main loops. In the second and third 131 Å images in Figure 3 (panels h and m), the flare arch is directed north-south and consists of a narrow pair of loops, one longer than the other. They appear to arch low into the corona and cross one above the other near the top of the arch. In the third 131 Å image (panel m), the longer flare loop (pointed to by the longer red arrow) is brighter and thicker than the shorter loop, perhaps arches above the shorter loop, and is oriented slightly clockwise to the shorter loop. In the fourth and fifth 131 Å images in Figure 2 (panels r and w), the flare-loop configuration has evolved to its final form. Now there are still basically two main loops that arch north-south, one shorter than the other, but instead of crossing each other they are parallel, side-by-side. The shorter loop is west of the longer loop, and is the brighter of the two.

Figure **6** is the schematic for eruption 2. Top left in Figure **6** is the proposed pre-reconnection magnetic field configuration viewed vertically from above, and top right is that field configuration viewed horizontally from the west. Bottom left and bottom right are the same two views of the north-south magnetic arch from the big negative-polarity sunspot and its side-by-side pair of heated flare loops at the end of the burst of reconnection.

From the 131 Å images in Figure 3 (panels h, m, r, w), we surmise that the burst of reconnection is in the body of the magnetic arch from the big sunspot and suddenly starts between two slightly askew crossed strands of the magnetic arch, at their interface where they cross and press against each other. We suppose that the longer of the two pre-reconnection magnetic strands arches over the shorter strand, so that in progressing along the positive-to-negative-flux direction of its magnetic field the longer strand crosses the shorter strand from east to west (as in the top left panel of Figure **6**) and presses against the upper west side of the shorter strand at the interface of the two strands (as in the top two panels of Figure **6**). This pre-reconnection setup is appropriate for a burst of reconnection at the interface to make and heat the two co-produced side-by-side flare loops, with the western loop being shorter and brighter than the eastern loop (as in the bottom two panels of Figure **6**). (The shorter loop is plausibly brighter than the longer loop if the burst of reconnection making the two loops gives each the same amount of heating energy.)

### *4.2. **Eruption** 1*

Eruption 1, shown in Figure 5, is like the other Table 1 flare eruptions in being a confined eruption. But eruption 1 differs from the other Table 1 flare eruptions in having for its final flare-loop configuration two crossed main loops instead of two side-by-side main loops.

In the third 131 Å image (panel m) in Figure 5, the flare arch is directed east-west and consists of a narrow pair of slightly skewed crossed loops – one longer than the other – that arch into the low corona. The longer loop (pointed to by the downward red arrow) is thinner and fainter than the shorter loop (pointed to by the upward red arrow), and apparently crosses above the shorter loop toward the west end of the shorter loop. In the fourth 131 Å image (panel r) in Figure 5, the feet of each of the two main loops are each centered about where they were in the second and third 131 Å images (panels h and m) in Figure 5, the body of each main loop has further brightened, and the shorter loop is now not as much brighter than the longer loop as it was in the third 131 Å image (panel m).

Figure 7 is the schematic for eruption 1. The proposed pre-eruption magnetic field configuration viewed vertically from above is in the top left panel of Figure **7**, and that field configuration viewed horizontally from the south is in the top right panel. The bottom left panel and bottom right panel are the same two views of the east-west magnetic arch from the big negative-polarity sunspot and its two crossed final main heated flare loops at the end of reconnection.

For the eruption 1 final flare-loop configuration to be the crossed pair of flare loops seen in Figure 5, we again surmise – as we have for eruption 2 – that the burst of reconnection is in the body of the magnetic arch from the big sunspot, but this time the reconnection is between two of the arch's magnetic strands – one longer than the other – that cross each other twice when viewed vertically from above, as in the top left panel of Figure 7. In going along the positive-to-negative-flux direction of its magnetic field, the pre-reconnection longer strand first crosses below the shorter strand from south to north and then crosses above the shorter strand from north to south toward the western end of the shorter strand. The two pre-reconnection strands press against each other only at their interface at their eastern crossing; they do not have an interface at their western crossing, as is depicted in the top right panel of Figure 7. This pre-reconnection setup is appropriate for a burst of reconnection at the eastern-crossing interface to make and heat in eruption 1 the two co-produced crossed flare loops, with the longer and less bright loop crossing from north to south over the shorter and brighter loop, as in the bottom two panels of Figure 7. (The shorter flare loop is plausibly brighter than the longer flare loop if the burst of reconnection that makes them puts the same amount of heating energy into each.)

## 5. Summary and Discussion

Using Helioviewer, we searched three years (2012, 2013, 2014) of full-disk AIA images and HMI magnetograms for flare eruptions that each occur in an ostensibly inert magnetic arch from a big sunspot. Our purpose was to find for examination with the good spatial and temporal resolution of AIA and HMI a sample of flare eruptions that are evidently similar to the umbral flares reported by Tang (1978). Our search yielded the seven flare eruptions listed and summarized in Table 1.

Each flare eruption evidently does not fit the standard model for single-bipole confined flare eruptions shown in Figure 1. Either the PIL of the pre-eruption magnetic arch is not sharp in the magnetograms and is not traced by a filament, or, if the PIL is traced by a filament, the filament is hardly disturbed by the eruption. Furthermore, the two flare ribbons of the flare arcade in the flaring magnetic arch start far from the PIL and do not noticeably spread apart as they grow in area and brightness.

Among our seven flare eruptions, the final configuration of main flare loops is two side-by-side loops in six eruptions, and is two crossed loops in one eruption. We present in detail two example eruptions: a representative example (eruption 2) of the flare eruptions that make two side-by-side main flare loops, and the flare eruption (eruption 1) that makes two crossed main flare loops. The six flare eruptions that make two side-by-side flare loops and the flare eruption that makes two crossed flare loops are each evidently a confined eruption contained within its enveloping magnetic arch from a sunspot.

For each of the two example flare eruptions, we present schematic drawings depicting the inferred configuration of the pre-eruption magnetic field and the field's reconnection that yields the final flare-loop configuration. The drawings for each example eruption illustrate how a burst of reconnection could make the onset and evolution of the observed flare loops. For the flare eruptions that make two side-by-side main flare loops, the burst of reconnection is evidently high inside the pre-eruption magnetic arch, plausibly at the interface between two high-arching strands of the magnetic arch that cross each other at a small (10° - 20°) angle where they press against each other near their tops. For the flare eruption that makes two crossed main flare loops, the burst of reconnection is again evidently high inside the pre-eruption magnetic arch. This time the reconnection is plausibly at an interface between two entwined high-arching strands of the magnetic arch that twist around each other at a small (10° - 20°) pitch angle, cross each other twice along the arch, and press against each other and reconnect at only one of the two crossings.

The schematic drawings in Figure **6** and Figure **7** display pre-eruption entwined pairs of magnetic strands (1) that have an interface where they cross each other high in the magnetic arch in which the eruption happens, and (2) for which a burst of reconnection at the interface plausibly makes two side-by-side main flare loops (in the case of Figure **6**) or two crossed main flare loops (in the case of Figure **7**). The drawings leave three major questions unanswered: (1) how do the two strands that suddenly reconnect in the flare eruption come to be entwined and press against each other before their reconnection, (2) why is the pair of entwined strands metastable against reconnection before their reconnection starts, and (3) what triggers the burst of reconnection? A plausible answer to the first question is that the two pre-reconnection strands become entwined by convection flows acting on them during and after the formation of the magnetic arch from the big sunspot.

Because the two main flare loops are close together and nearly parallel in all six eruptions having two side-by-side main flare loops, and cross each other at a small (10° - 20°) angle in the one eruption having two crossed main flare loops, we infer that at the start of their reconnection the entwined pair of strands cross each other at a small (10° - 20°) angle. That suggests the answers to the second and third question are: (2) the interface is stable against reconnection when the crossing angle is small enough, and (3) reconnection suddenly starts at the interface when convection-driven evolution of the magnetic arch increases the crossing angle at the interface to a critical (large enough) angle of 10° - 20°.

Parker (1983, 1988) proposed that the heating of the coronal plasma in coronal magnetic loops such as in solar active regions is by nanoflare bursts of reconnection at interfaces of pairs of a magnetic loop's substrands that have been entwined by random braiding by the convection flows of the photospheric granulation at the loop feet, and that a nanoflare burst of reconnection at an interface between two entwined strands occurs when the crossing angle of the two strands at the interface grows to a critical (large enough) angle. Parker (1988) estimated the critical crossing angle is $\sim$ 14°. Our observations of non-standard-model confined flare eruptions in magnetic arches from big sunspots suggest such non-standard-model confined flare eruptions result when two crossed strands in the arch suddenly undergo a burst of reconnection at their interface, and that the burst of reconnection is triggered when the crossing angle at the interface of the two strands grows to 10° - 20°; that is, grows to about the magnitude of the critical angle estimated by Parker (1988) for nanoflares to provide the heating of the coronal plasma in active-region coronal loops. From the implication of our observations that the critical crossing angle for triggering a burst of reconnection is about that estimated by Parker (1988) for nanoflares in active-region coronal loops, we judge that our non-standard-model confined flare eruptions in magnetic arches from big sunspots support Parker's basic idea that the coronal heating in solar coronal loops is by nanoflare bursts of reconnection at magnetic-strand interfaces made in coronal magnetic loops by braiding of the strands by the magnetoconvection at their feet. (Parker's idea is also empirically supported by the Tiwari et al 2017 observation that active-region coronal loops having opposite feet in umbrae of opposite-polarity sunspots in bipolar active regions are dark in coronal-temperature EUV emission, presumably because they lack coronal heating due to magnetoconvection in sunspot umbrae being strongly suppressed by the strong magnetic field there.)

Further, from the agreement between our critical-angle estimate for the bursts of reconnection for our non-standard-model confined flares and Parker's critical-angle estimate for nanoflare bursts of reconnection, we posit that the braiding of active-region coronal loop magnetic strands by magnetoconvection at the loop feet – perhaps in combination with the helicity-condensation process proposed by Antiochos (2013) – results in an occurrence-rate spectrum of non-standard-model confined flares that decreases steeply with increasing event energy. The spectrum is populated near its small-event-energy end by Parker's nanoflares, and toward its large-event-energy end by flare events of the magnitude of the seven non-standard-model confined flare events in Table 1 and bigger (up to at least the magnitude of the GOES M1 non-standard-model confined flare presented by Tandberg-Hanssen et al 1984).

We judge that the type of non-standard-model confined flare reported here is importantly different from either of two previously reported types of non-standard-model confined flare. One previously reported type is that of the loop-loop-reconnection flares reported by Nishio et al (1997) and Hanaoka (1997). These flare eruptions evidently result from a burst of reconnection at the interface between an emerging or recently emerged short magnetic loop and the encountered leg of a relatively long magnetic loop. It is not clear from the observations presented by Nishio et al (1997) and Hanaoka (1997) that the angle between the short loop and the long loop at their interface is as small as the 10° - 20° angle that we infer for our flare eruptions. In any case, the inferred magnetic strands that reconnect in our eruptions are much more nearly the same length than the two loops that reconnect in the flares reported by Nishio et al (1997) and Hanaoka (1997), and there is no indication that the inferred interface for reconnection in our eruptions results from an emerging magnetic loop encountering another magnetic loop. The other previously reported type of non-standard-model confined flare eruption is the type called "Type I" by Li et al (2019). Confined flare eruptions of this type occur in obviously strongly sheared magnetic field that envelops a PIL that is typically traced by filament. During a "Type I" confined flare, the filament remains stationary. Even so, "Type I" confined flare eruptions strikingly differ from our flare eruptions in that for "type 1" confined flare eruptions (1) the pre-flare magnetic field is ostensibly capable of flaring because it is obviously strongly sheared and hence obviously has a large store of free magnetic energy, instead of being ostensibly not capable of flaring in the manner of our pre-flare magnetic arches, and (2) their flare ribbons rapidly spread in the direction along the PIL of the sheared magnetic field in which the flare erupts, whereas there is no such rapid ribbon spreading in our confined flare eruptions.

Due to the above differences, even though the flare energy is presumably released by reconnection in closed magnetic field in each type of non-standard-model confined flare eruption, we judge that our type of non-standard-model confined flare eruption is significantly different from the above two previously reported types. Especially, we judge that our non-standard-model confined flare eruptions are more suggestive of – and hence more supportive of – the Parker (1983, 1988) idea for coronal loop heating by nanoflaring than are the non-standard-model confined flares of either Nishio et al (1997) and (Hanaoka 1997) or Li et al (2019).

We thank the referee for carefully reading the paper and for providing constructive comments. This work was supported by the NASA Science Mission Directorate's Heliophysics Division by research grants from the Heliophysics Guest Investigators program and the Heliophysics Supporting Research program.

| Table 1. Observed Aspects of Seven Flare or Subflare Confined Eruptions in Ostensibly Inert Magnetic Arches from Sunspots | | | | | | | | | |
|---|---|---|---|---|---|---|---|---|---|
| Eruption Number | Date (YYYY MM DD) & Place (N/S, E/W Degrees) | Active Region (NOAA Number) | Flare Class ( Flare/ Subflare, GOES Magnitude) | Underlying PIL | | Heated Loops | | | Further Aspects |
| | | | | Sharp? (Yes/No) | Underlies Filament or Filament Chanel? (Yes/No) | Start (UT) | End (UT) | Configuration (Two Crossed Loops/ Two Side-by-Side Loops/) | |
| 1 | 2012 Jan 8 N12, E10 | AR 11391 | Subflare, <~ C1.0 | No | YES | 05:55 | 06:30 | Two Crossed Loops | Eruption doesn't disturb filament. Flare loop feet far from PIL. Conjugate feet don't spread apart. Shorter loop brighter. Longer loop wider. |
| 2 | 2013 Jan 14 N03, W09 | AR 11654 | Flare, C2.5 | No | No | 17:10 | 18:05 | Two Side-by-Side Loops (at end of heating) | Initial heated loops are a crossed pair. Feet far from PIL. Conjugate feet don't spread apart. Shorter loop brighter, cools sooner. |
| 3 | 2013 Jun 27 S13, W14 | AR 11777 | Subflare, B5.0 | No | No | 14:50 | 15:50 | Two Side-by-Side Loops | Feet far from PIL. Conjugate feet don't spread apart. Shorter loop brighter, cools sooner. Longer loop wider, fan-shaped. |
| 4 | 2014 Jan 23 S20, E09 | AR 11959 | Subflare, ~ B3 | No | Yes | 09:45 | 10:35 | Two Side-by-Side Loops | Eruption doesn't disturb filament. Flare loop feet far from PIL. Conjugate feet don't spread apart. Shorter loop brighter, cools sooner. Longer loop wider, fan-shaped. |
| 5 | 2014 Nov 19 S11, E02 | AR 12209 | Flare, C2.0 | No | No | 18:55 | 19:30 | Two Side-by-Side Loops | Feet far from PIL. Conjugate feet don't spread apart. Shorter loop brighter, cools sooner. Longer loop wider, fan-shaped. |
| 6 | 2014 Nov 19 S11, E02 | AR 12209 | Flare, C1.0 | No | No | 20:30 | 21:30 | Two Side-by-Side Loops | Feet far from PIL. Conjugate feet don't spread apart. Shorter loop brighter, cools sooner. Longer loop wider, fan-shaped. |
| 7 | 2014 Nov 20 S11, W09 | AR 12209 | Flare, C2.4 | No | No | 19:40 | 20:45 | Two Side-by-Side Loops | Feet far from PIL. Conjugate feet don't spread apart. Shorter loop brighter, cools faster. Longer loop wider. |

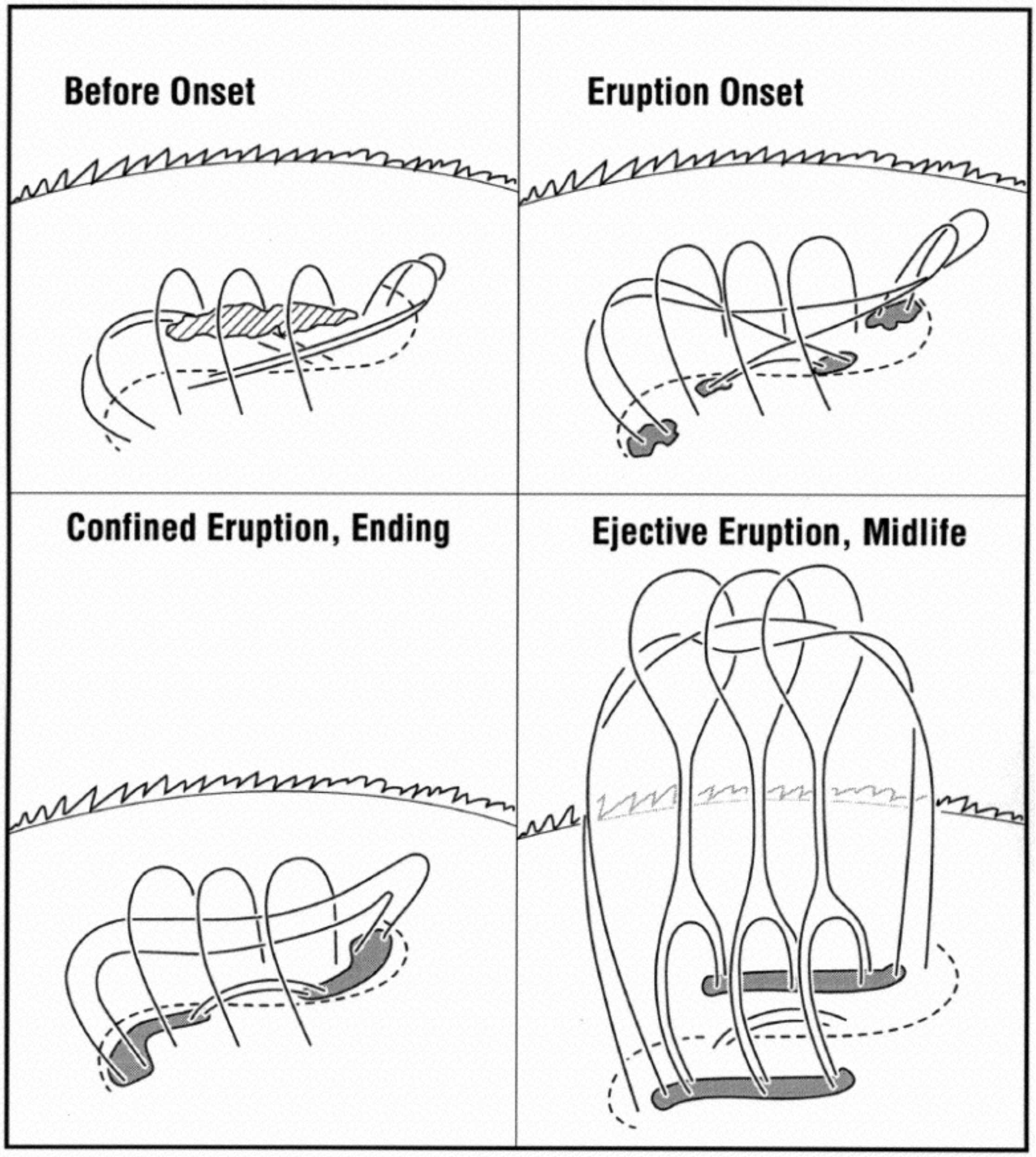

Figure 1. The Moore et al (2001) schematic of the standard model for confined flare eruptions and ejective flare eruptions in lone bipolar active regions. Solid curves are magnetic field lines. The dashed curve is the active region's polarity inversion line (PIL). The elongated shaded feature in the first panel is a filament of dark chromospheric-temperature plasma suspended in the active region's sheared core field above the PIL. Dark areas are flare ribbons at the feet of reconnected field lines. The background ragged arc is the chromospheric limb.

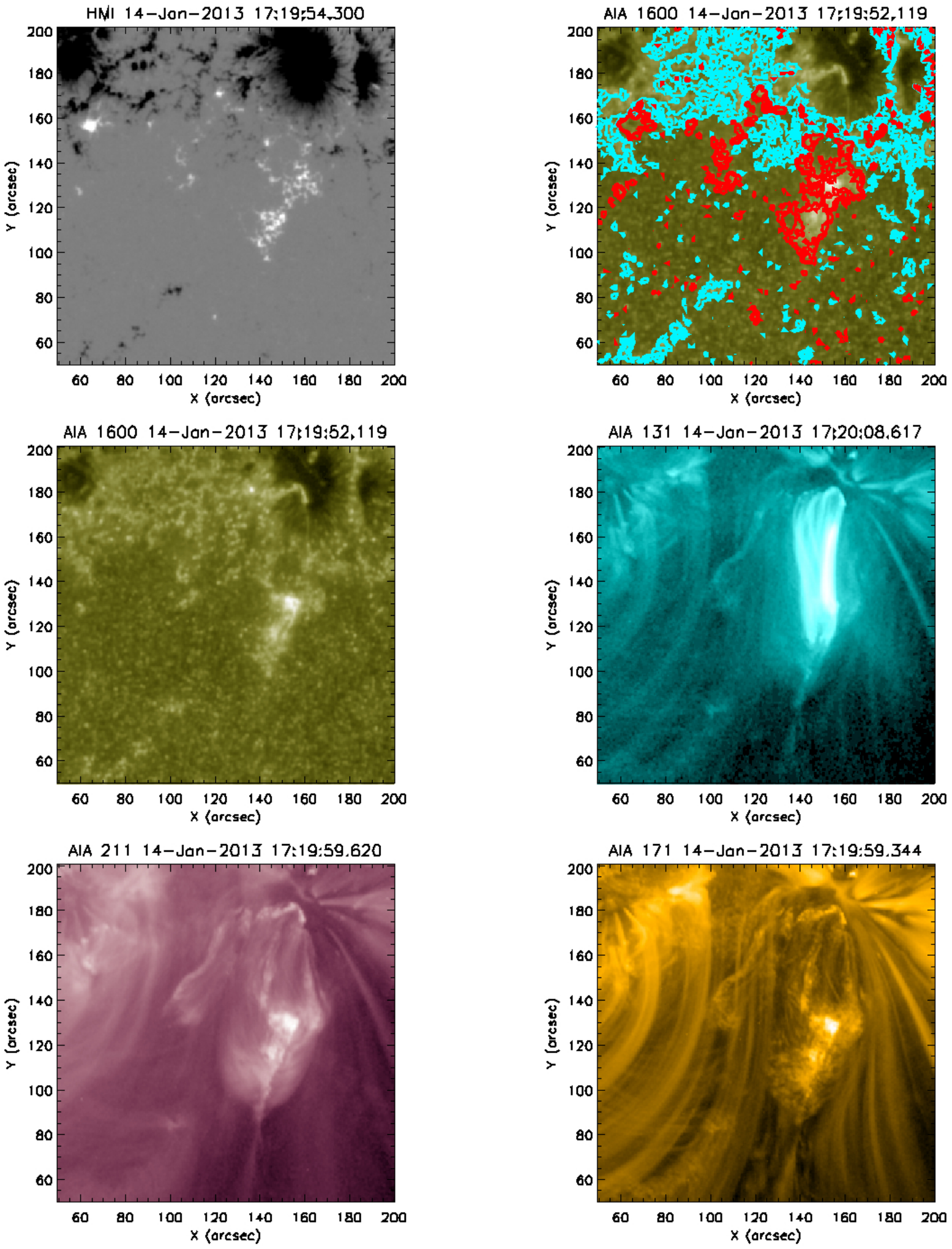

Figure 2. A single frame from **Figure 2 animation,** for **eruption** 2 of Table 1. Here and in all subsequent figures in this paper, heliographic north is up and west is to the right. The time of this frame is near the time of maximum luminosity of **eruption** 2 in AIA 131 Å emission. All six panels are at nearly the same time and cover the same heliographic area encompassing **eruption** 2. Top left: HMI magnetogram. Middle left: AIA 1600 Å image. Top right: magnetogram 20 G and 100 G contours (red for positive flux, blue for negative flux) overlaid on the 1600 Å image. Middle right: AIA 131 Å image. Bottom left: AIA 211 Å image. Bottom right: AIA 171 Å image. The 131 Å image shows the final configuration of the two main flare loops, and the 1600 Å, 211 Å, and 171 Å images show the flare ribbons at the feet of the flare loops. The top right panel shows that the northern feet of the flare loops are in negative magnetic flux and the southern feet are in positive flux. The animation spans the 55-minute duration given for **eruption** 2 in Table 1. The cadence of the magnetograms is 45 s. The cadence of the 1600 Å images is 24 s. The cadence of the 131 Å, 211 Å, and 171 Å images is 12 s. **The animation starts at 17:09 UT and ends at 18:03 UT on 2013 January 14, and has a duration of about 16 seconds.**

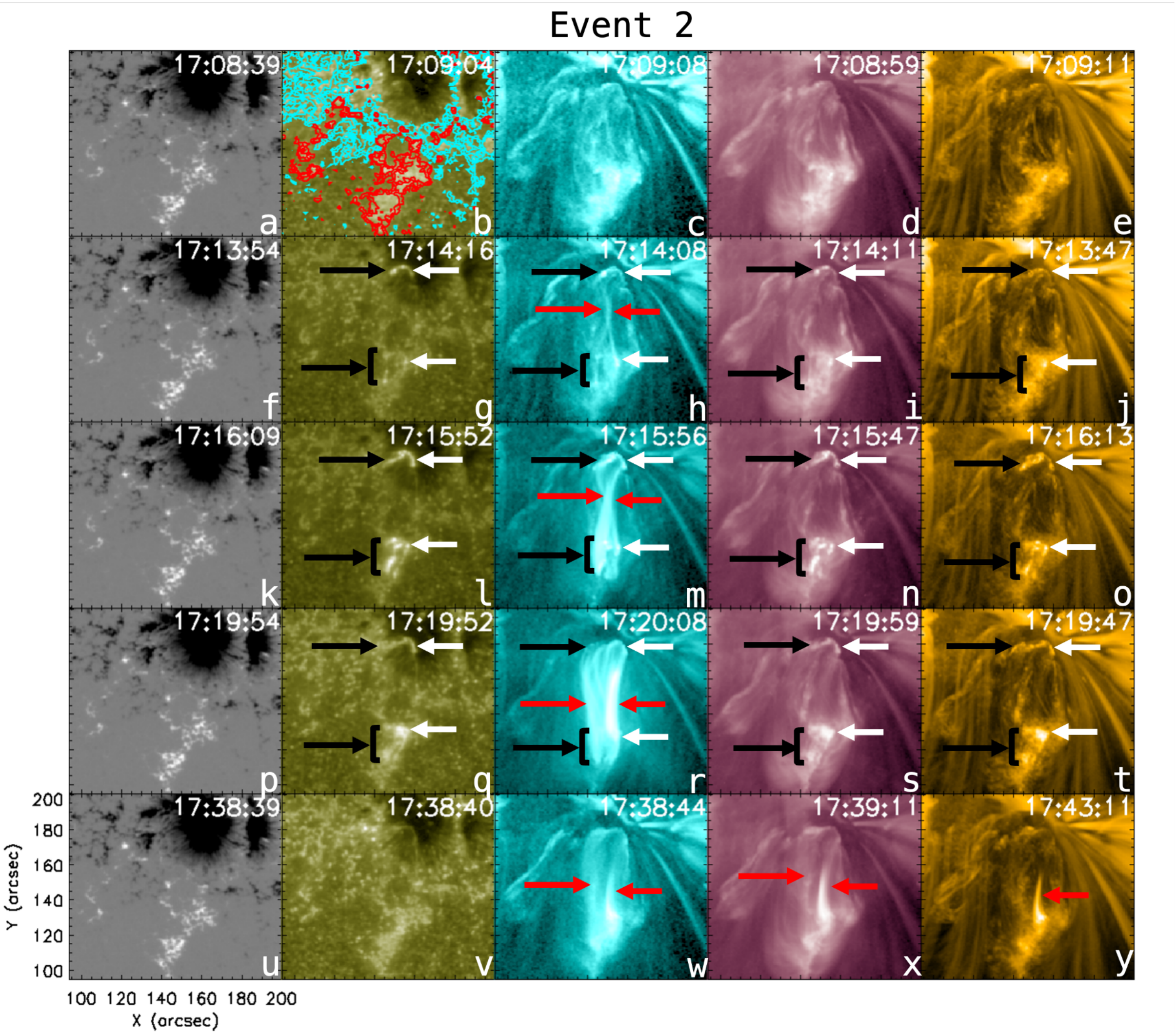

Figure 3. Magnetic setting and progression of the flare ribbons and loops in **eruption** 2 of Table 1. First column: HMI magnetograms; second column: AIA 1600 Å images; third column: AIA 131 Å images; fourth column: AA 211 Å images; fifth column: AIA 171 Å images. The x, y coordinates of the field of view in arc seconds from disk center are on the outside of the last magnetogram. The universal time is in the upper right corner of each panel. In the first row, 20 G and 100 G contours (red for positive flux, blue for negative flux) of the magnetogram are plotted on the 1600 Å image. Black arrows and white arrows point to the four feet of the two main flare loops. In panels r, w, and x the long red arrow points to the longer and dimmer of the two loops, and the short red arrow points to the shorter and brighter of the two loops.

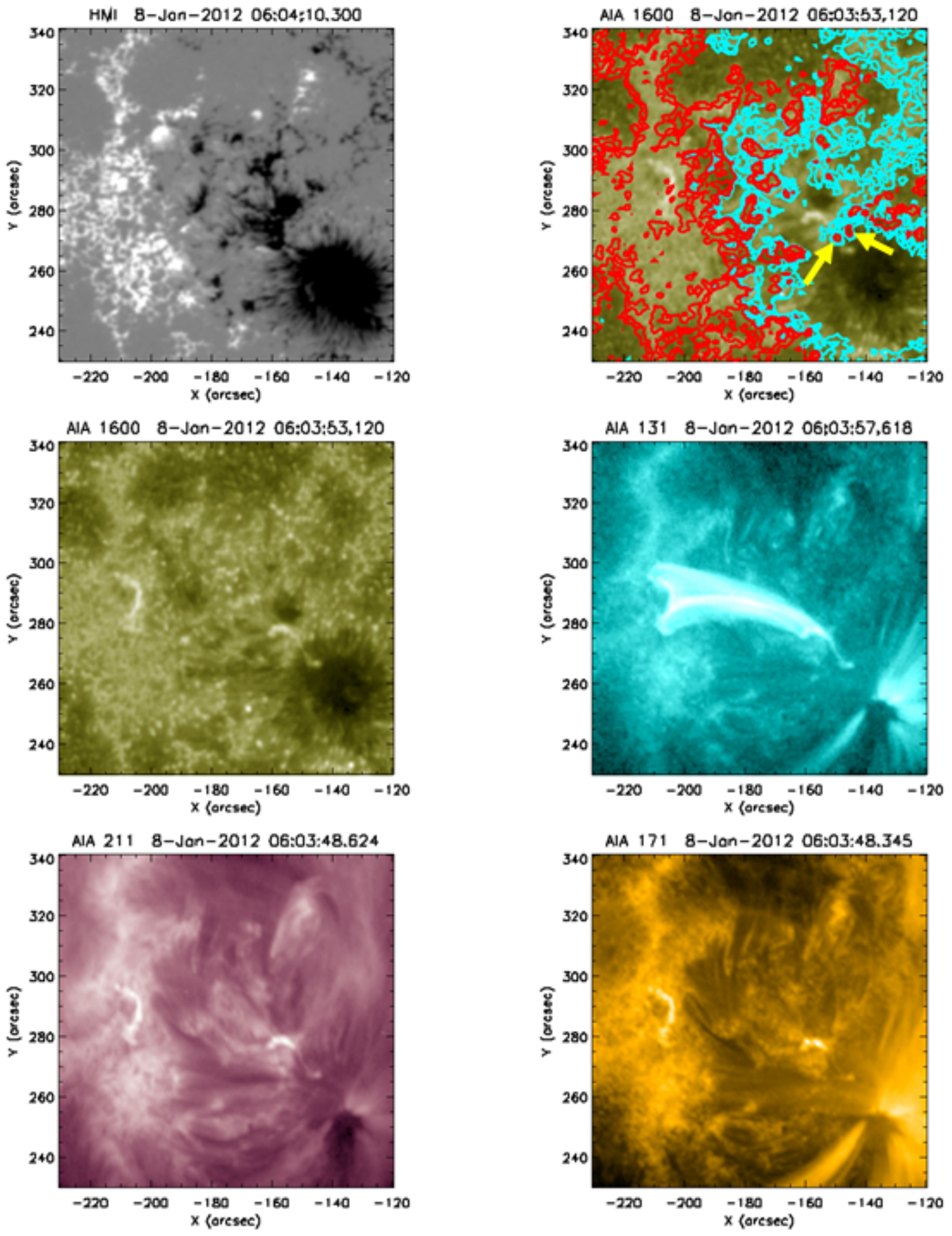

Figure 4. A single frame from **Figure 4 animation,** for **eruption** 1 of Table 1. The format is the same as in Figure 2. The time of this frame is near the time of maximum luminosity of **eruption** 1 in AIA 131 Å emission. The 131 Å image shows the final configuration of the two main flare loops, and the 1600 Å, 131 Å, 211 Å, and 171 Å images show the flare ribbons at the feet of the flare loops. The top right panel shows that the eastern feet of the flare loops are in positive magnetic flux, that the western feet of the flare loops are at least mostly in negative flux, and that some of the western feet are rooted near two tiny islands of positive flux. Each of these two islands of positive flux is pointed to by an arrow. The animation spans the 35-minute duration given for Event 1 in Table 1. The cadence is 45 s for the magnetograms, 24 s for the 1600 Å images, and 12 s for the 131 Å, 211 Å, and 171 Å images. **The animation starts at 05:54 UT and ends at 06:28 UT on 2012 January 8, and has a duration of about 11 seconds.**

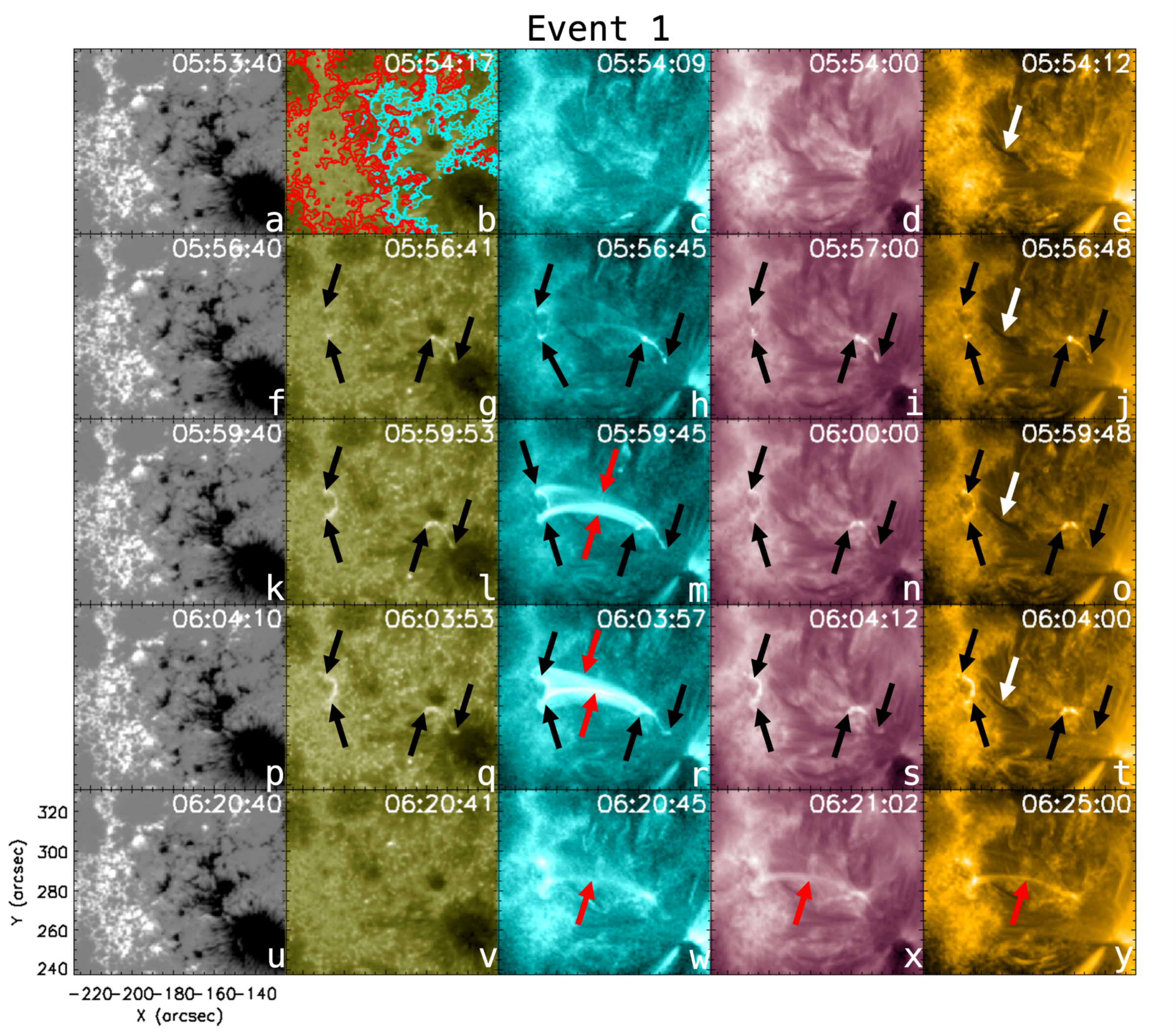

Figure 5. Magnetic setting and progression of the flare ribbons and flare loops in **eruption** 1 of Table 1. The format is the same as in Figure 3. Black arrows point to the four feet of the two main flare loops. Downward red arrows point to the longer and dimmer of the two loops. Upward red arrows point to the shorter and brighter of the two loops. In four of the of the 171 Å images, white arrows point to an inert dark filament suspended above a ragged PIL seen in the magnetogram.

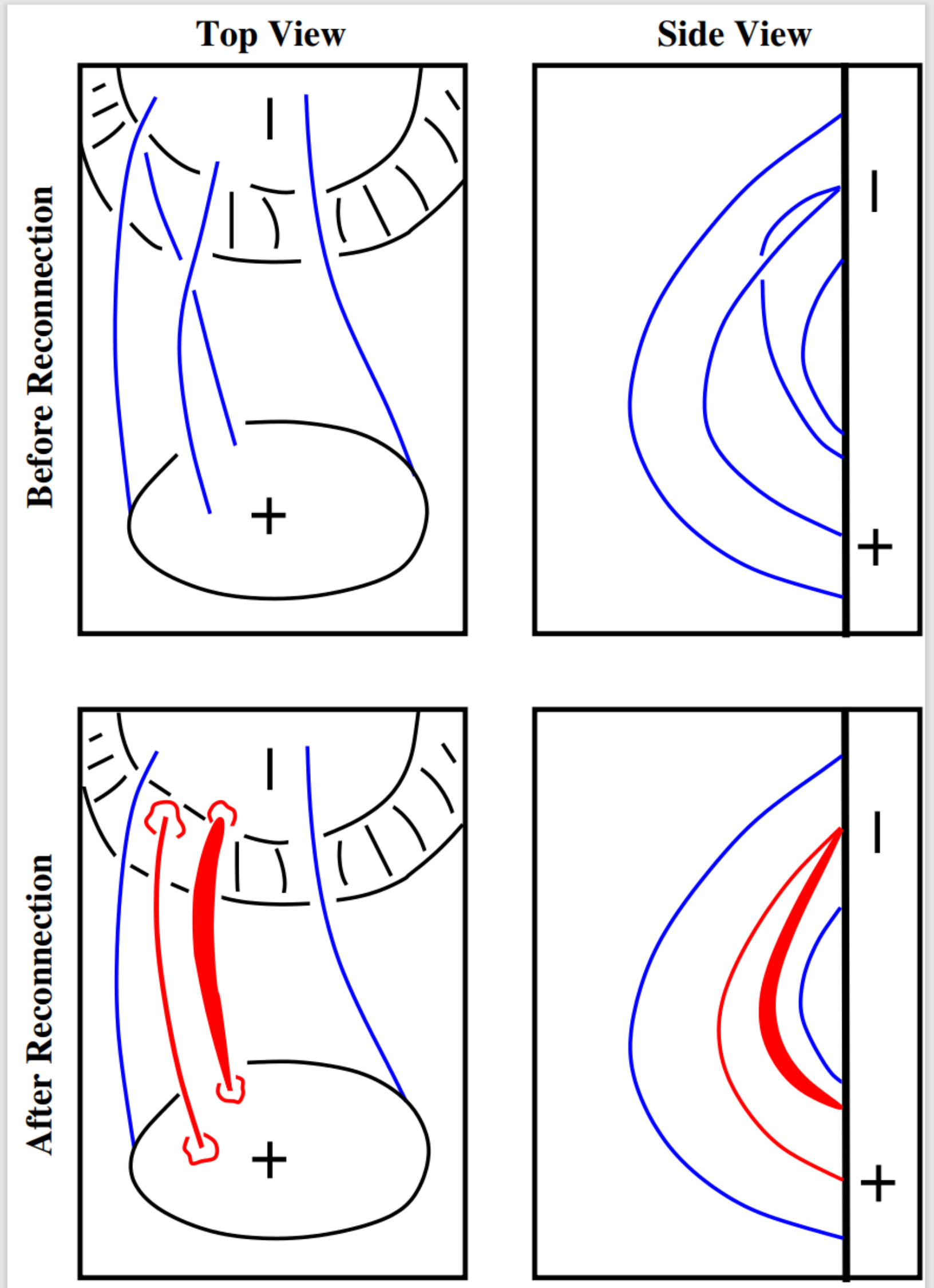

Figure 6. Schematic of the proposed internal structure of the magnetic arch from the big sunspot in AR 11654 before and after the burst of reconnection that makes and heats the two side-by-side main flare loops in **eruption** 2. Top: the magnetic arch viewed from above (left panel) and from its west side (right panel) before the burst of reconnection. Bottom: the same two views after the burst of reconnection. The minus (plus) signs indicate the negative (positive) polarity of the magnetic arch's end inside (outside) the sunspot. Black curves outline the magnetic arch's outer end and the sunspot's umbra and penumbra. The body of the magnetic arch is outlined by blue curves. In the interior of the magnetic arch before the burst of reconnection, the two blue curves are the two crossed magnetic field strands that are poised to reconnect at their interface. The two side-by-side main flare loops made by the burst of reconnection are red, and outlines of the flare "ribbons" at their feet are red.

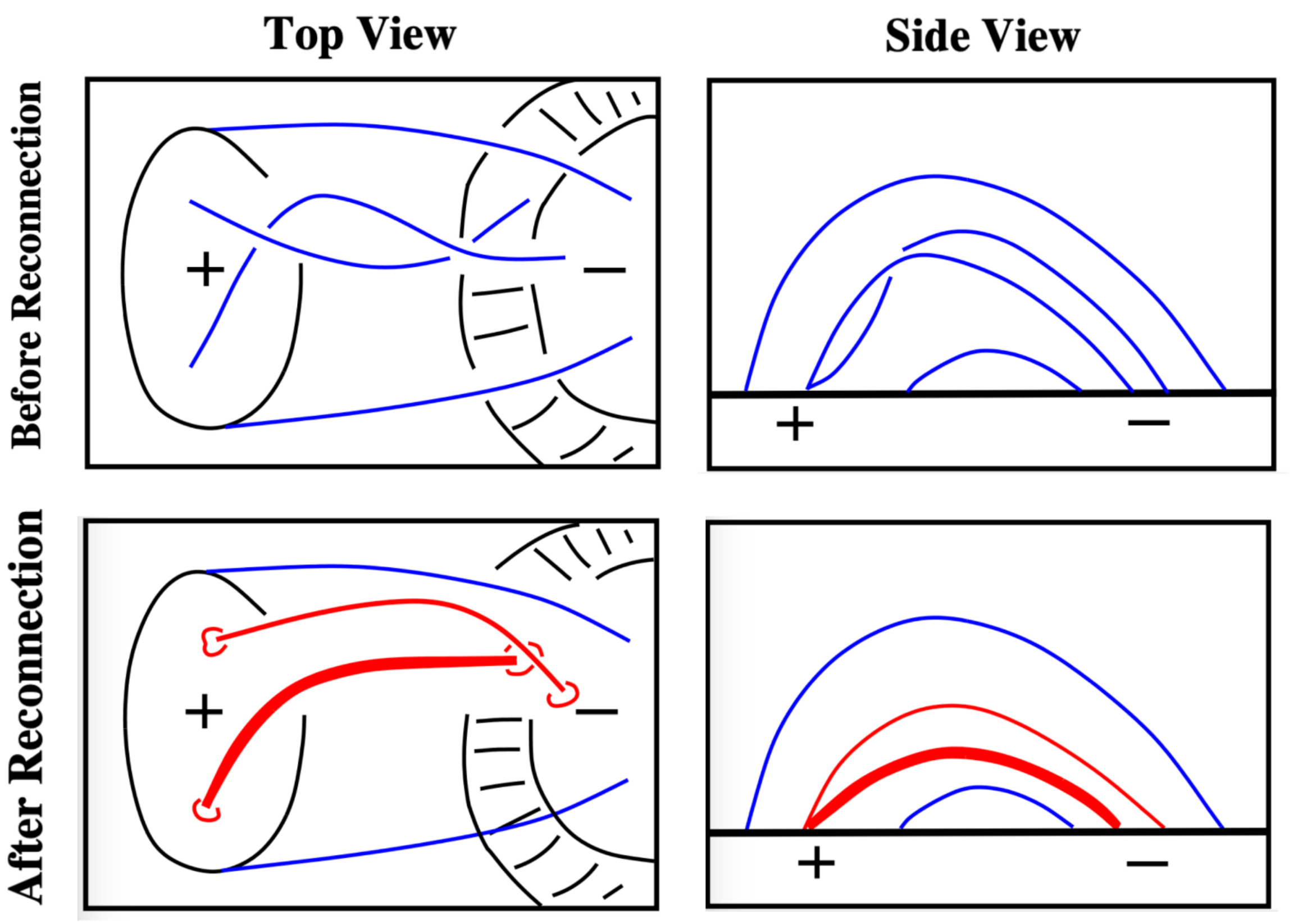

Figure 7. Schematic of the proposed internal structure of the magnetic arch from the big sunspot in AR 11391 before and after the burst of reconnection that makes and heats the two crossed main flare loops in **eruption** 1. The format and symbolism are the same as in Figure 6. The two pre-reconnection magnetic field strands cross twice in the top view but only once in the side view. That shows they are poised to reconnect at only their eastern crossing. The burst of reconnection at their interface at their eastern crossing results in the two flare loops, with the longer and dimmer flare loop crossing above the west end of the shorter and brighter flare loop.

**Animation** Legends:

**Figure 2 animation**. Top left panel: HMI magnetograms. Middle left panel: AIA 1600 Å images. Top right panel: AIA 171 Å images overlaid with 20 G and 100 G magnetogram contours (red for positive flux, blue for negative flux). Middle right panel: AIA 131 Å images. Bottom left panel: AIA 211 Å images. Bottom right panel: AIA 171 Å images. The **animation** spans the 55-minute duration given for eruption 2 in Table 1. The cadences are 45 s for the magnetograms, 24 s for the 1600 Å images, and 12 s for the 131 Å, 211 Å, and 171 Å images. The animation starts at 17:09 UT and ends at 18:03 UT on 2013 January 14, and has a duration of about 16 seconds.

**Figure 4 animation**. The format and cadences are the same as for Figure 2 animation. The animation spans the 35-minute duration given for eruption 1 in Table 1. The animation starts at 05:54 UT and ends at 06:28 UT on 2012 January 8, and has a duration of about 11 seconds.